\documentclass[mathleft,fleqn,%
]{an}
\usepackage{graphicx,amssymb,amsmath}
\usepackage[varg]{txfonts}
\usepackage{natbib}
\bibpunct{(}{)}{;}{a}{}{,}
\newcommand{\cm}{\,\mathrm{cm}}
\newcommand{\FRM}{\,\mathrm{rad\,m^{-2}}} 
\newcommand{\kpc}{\,\mathrm{kpc}}
\newcommand{\pc}{\,\mathrm{pc}}

\newcommand{\uJyb}{\,\mu\mathrm{Jy/beam}} 

\newcommand{\removed}[1]{}  

\newcommand{\infint}{\int_{-\infty}^{\infty}}
\newcommand{\wwtmax}{w^\mathrm{max}}
\newcommand{\wmaxwt}{\tilde{w}}

\begin{document}

\Pagespan{1}{}
\Yearpublication{2018}%
\Yearsubmission{2018}%
\Month{0}%
\Volume{999}%
\Issue{0}%
\DOI{asna.201400000}%

\title{Magnetic arms of NGC6946 traced in the Faraday cubes at low radio frequencies}
\author
{A. Chupin$^{15}$\thanks{E-mail: chupin@icmm.ru}, R. Beck$^2$, P. Frick$^1$, G. Heald$^3$, D. Sokoloff$^4$, R. Stepanov$^1$}

\titlerunning{Magnetic arms traced in the Faraday cube}
\authorrunning{A. Chupin et al.}
\institute{
Institute of Continuous Media Mechanics, Korolyov str. 1, 614061 Perm, Russia
\and 
MPI f\"{u}r Radioastronomie, Auf dem H\"{u}gel 69, 53121 Bonn, Germany
\and 
CSIRO Astronomy and Space Science, 26 Dick Perry Avenue, Kensington, WA 6151, Australia; \\
ASTRON, the Netherlands Institute for Radio Astronomy, Postbus 2, NL-7990 AA Dwingeloo, the Netherlands; \\
Kapteyn Astronomical Institute, University of Groningen, PO Box 800, NL-9700 AV Groningen, the Netherlands
\and
Department of Physics, Moscow University, 119992 Moscow, Russia
\and
Perm State National Research University, Perm, Russia
}

\received{XXXX}
\accepted{XXXX}
\publonline{XXXX}


\abstract{%
Magnetic fields in galaxies exist on various spatial scales. Large-scale magnetic fields are thought to be generated by the $\alpha-\Omega$ dynamo. Small-scale galactic magnetic fields (1\,$\kpc$ and below) can be generated by tangling the large-scale field or by the small-scale turbulent dynamo. The analysis of field structures with the help of polarized radio continuum emission is hampered by the effect of Faraday dispersion (due to fluctuations in magnetic field and/or thermal electron density) that shifts signals from large to small scales. At long observation wavelengths large-scale magnetic fields may become invisible, as in the case of spectro-polarimetric data cube of the spiral galaxy NGC~6946 observed with the Westerbork Radio Synthesis Telescope in the wavelength range 17--23\,cm. The application of RM Synthesis alone does not overcome this problem. We propose to decompose the Faraday data cube into data cubes at different spatial scales by a wavelet transform. Signatures of the ``magnetic arms'' observed in NGC~6946 at shorter wavelengths become visible. Our method allows us to search for large-scale field patterns in data cubes at long wavelengths, as provided by new-generation radio telescopes.
}

\maketitle

\section{Introduction}

Magnetic fields of nearby galaxies are quite well investigated. The observational results are compatible with a scenario of magnetic field excitation by a galactic $\alpha-\Omega$ dynamo \citep[for a review see e.g.][]{Betal96}. The main bulk of these observations is obtained from analysis of galactic polarized radio continuum radiation observed with the current generation of radio telescopes, such as the Effelsberg 100-m dish and the Very Large Array (VLA).

Galactic magnetic fields are important for the evolution of galaxies, the astrophysics of the interstellar medium, cosmic ray propagation \citep[e.g.][]{Pr}, as well as of fundamental interest. In particular, observations demonstrate prominent magnetic structures in the form of magnetic arms that are usually situated between material arms, as in NGC~6946 \citep{2007A&A...470..539B}. The relation between gas and magnetic fields can also be more complicated, like e.g. in IC~342 \citep{Beck2015} and in M\,83 \citep{Frick16}. The verification of various scenarios for the generation of such fine magnetic structures \citep[e.g.][]{Metal15,Anvetal13} is limited by technical abilities (angular resolution and sensitivity) of current radio telescopes. Further progress can be expected with the new generation of radio telescopes, the European Low Frequency Array (LOFAR), the Australia SKA Pathfinder (ASKAP) and the South-African Karoo Array Telescope (MeerKAT), allowing high-resolution, multichannel polarimetric observations.

An important effect for the quantification of magnetic fields is Faraday rotation of polarized radio radiation that requires multi-wavelength observations \cite[e.g.][]{RS79}, while contemporary models for galactic magnetic fields are based on observations on few (and quite often of two) wavelengths only. As the Faraday rotation angle increases with the square of wavelength, observing at substantially longer wavelengths compared to those ones for the Effelsberg and VLA telescopes (3--20\,cm) increases the accuracy of the measurements substantially \citep{Beck2012}.
On the other hand, increasing the observation wavelength leads to more severe Faraday depolarization effects, which
complicates the extraction of physically valuable information on magnetic fields from observational data \citep{B66,Setal98}.

\begin{figure}
\centering
    \includegraphics[width=0.35\textwidth]{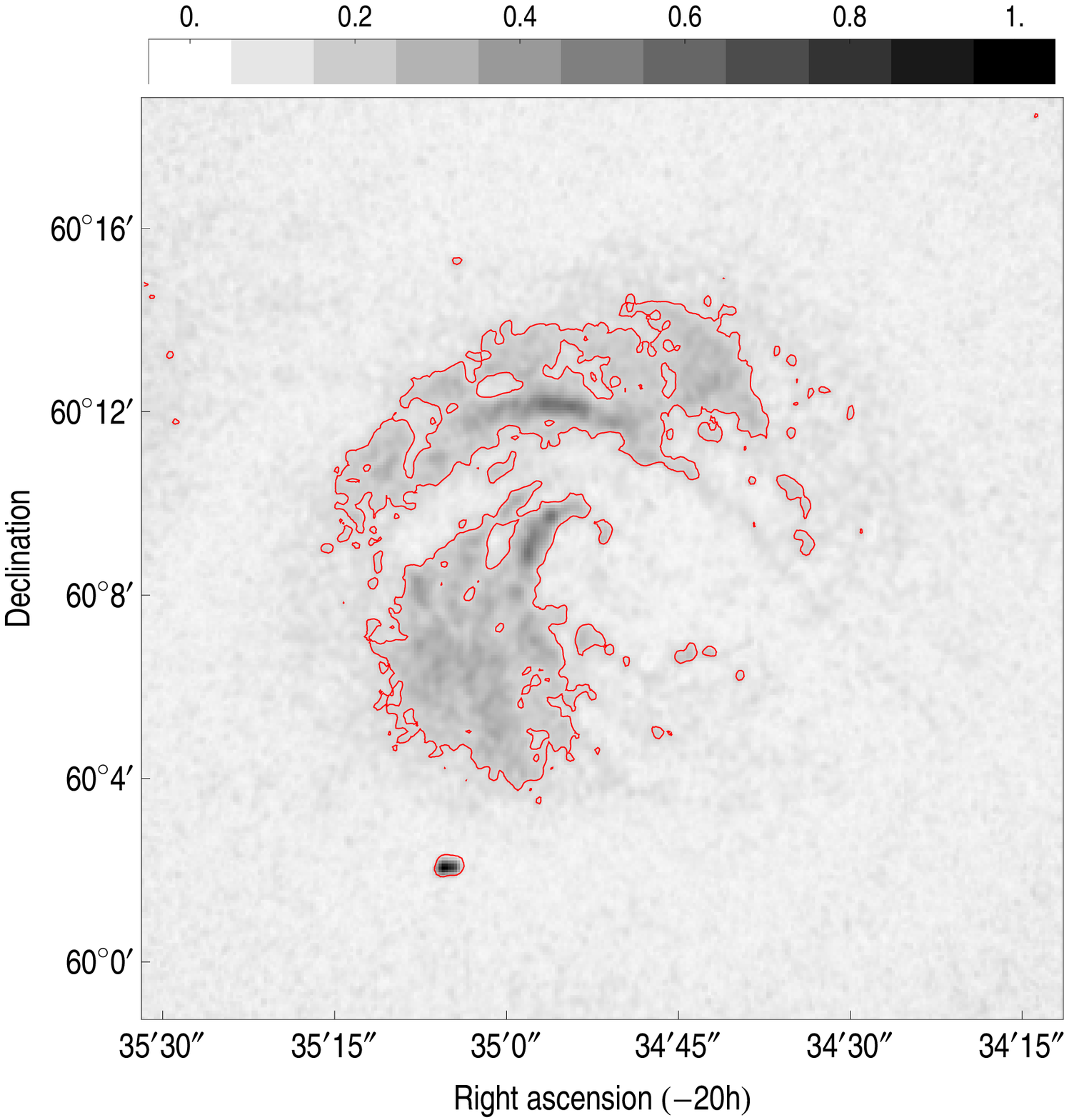}
    \includegraphics[width=0.35\textwidth]{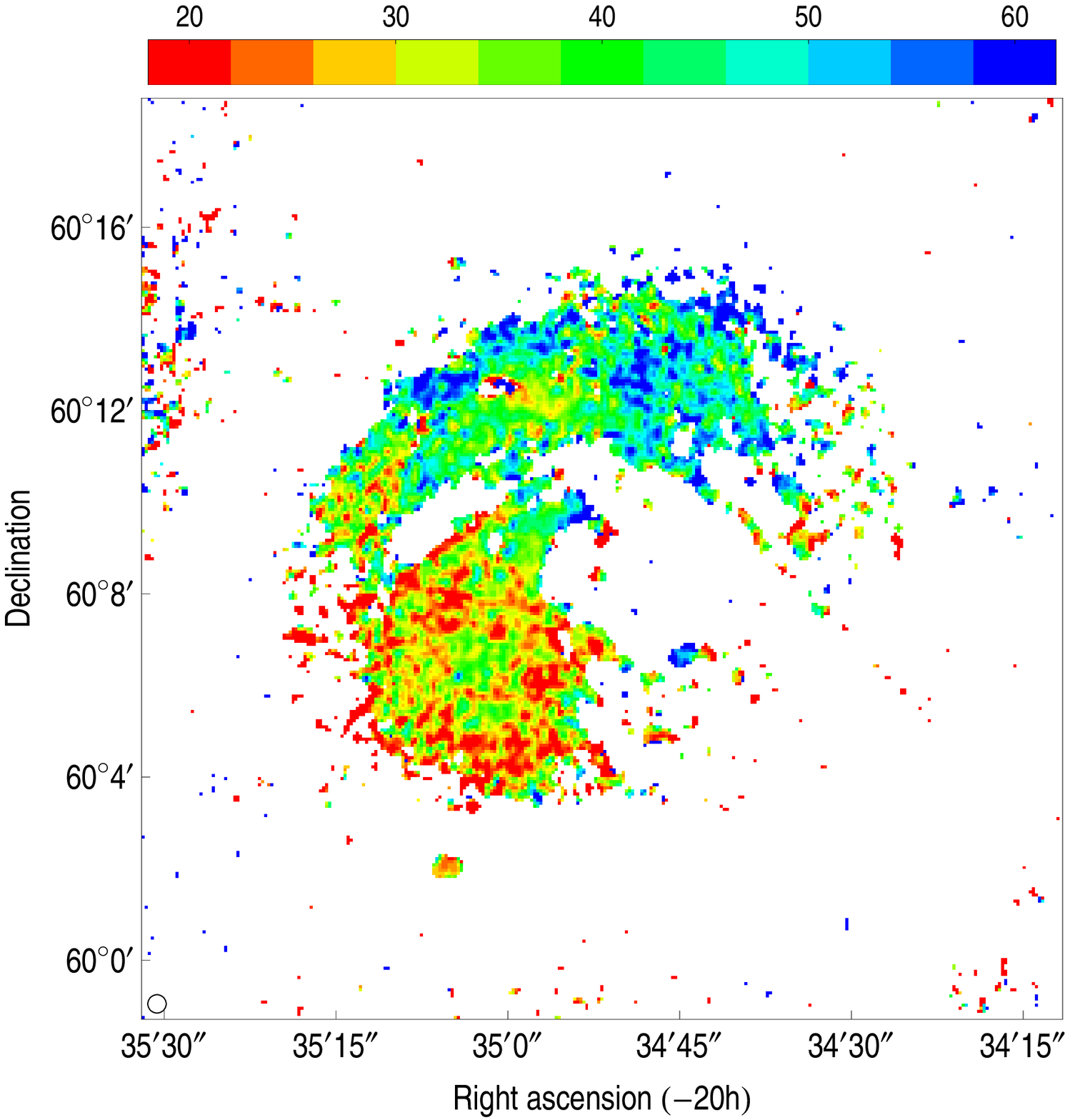}
\caption{Observations of NGC~6946 in the wavelength range 17--23\,cm and results of RM Synthesis:
top - distribution of a (normalized) peak intensity $|F^{\rm max}|$ of the Faraday spectra with overlayed red contours at 15\% of the maximum,
bottom - distribution of Faraday depths $\phi^{\rm max}$ (in $\FRM$) at which the maximal values of $|F^{\rm max}|$ are obtained, according to \citet{Heald2009}.}
\label{fig:data}
\end{figure}

Faraday depolarization, in particular Faraday dispersion due to small-scale fluctuations in magnetic field and thermal electron density within the emitting medium or in the Faraday-rotating medium in the foreground \citep{Setal98}, can already be strong at wavelengths of around 20\,cm. Maps of polarized radio emission obtained with the Westerbork Synthesis Radio Telescope (WSRT) for many (21) galaxies \citep{Heald2009} do not reveal (at least not in a straightforward way) many structures that are well known from observations at shorter wavelengths.
In particular, the map of Faraday depths (a measure of the integral of the product of thermal electron density $n_\mathrm{e}$ and magnetic field strength $B$ along the line of sight) of NGC~6946 obtained by \citet{Heald2009}
in the wavelength range 17--23\,cm (consisting of two bands of 17.0--18.4\,cm and 20.9--23.1\,cm) (Fig.~\ref{fig:data}, bottom)
 shows small-scale fluctuations superimposed on a large-scale gradient
and does not directly show the well-defined ``magnetic arms'' known from a previous study at $\lambda$ 3.6\,cm and 6.3\,cm by \citet{2007A&A...470..539B}. Fig.~\ref{fig:data} was obtained using the sophisticated method of RM Synthesis \citep{Brentjens2005}, but this does not overcome this problem.

The aim of this paper is to demonstrate that by combining the ideas of wavelet analysis and RM Synthesis one can recover the magnetic arm configuration from observations at long radio wavelengths.

The key idea of the approach is as follows. Strong Faraday depolarization randomizes almost all information concerning the large-scale magnetic field structure. Faraday rotation angles $0.81 \, B \, n_e \, d \, \lambda^2$ (where $B$ is the average strength of the magnetic field along the line of sight, $d$ is the coherence scale of the magnetic field, $n_\mathrm{e}$ is electron density and $\lambda$ the wavelength) are generally larger than $\pi$ at $\lambda\simeq20$\,cm. Fluctuations in $B$ and $n_\mathrm{e}$ lead to strong gradients in the maps of Stokes Q and U and hence to shifting the signals of polarized intensity from large to small angular scales. The power spectrum of polarized intensity becomes flatter \citep[e.g.][]{2006A&A...457....1L}.
The imprints of the large-scale field remain recognizable at smaller spatial scales. By using wavelets we can isolate small-scale magnetic fields obtained as the result of decay of the large-scale ones and then recognize the locations of large-scale fields.

We will test our method on the observational data for NGC~6946 by \citet{Heald2009}. The magnetic arm configuration in this galaxy is known from observations at short wavelengths \citep{2007A&A...470..539B}, allowing us to verify the results.

\section{Method and results}

We base our analysis on observation of linearly polarized radio continuum emission of NGC~6946 in the range 17--23\,cm obtained as part of the Spitzer Infrared Nearby Galaxies Survey (SINGS) by the Westerbork Synthesis Radio Telescope (WSRT) with an angular resolution (beam size) of $15\arcsec \times 17\arcsec$ \citep{Heald2009} (Fig.~\ref{fig:data}).
The data is available in the form of
cubes of Stokes parameters Q and U in 803 frequency channels. The scale is 4\arcsec\ per pixel. Images at 11\,cm from Effelsberg and at 6\,cm combined from data from the VLA and Effelsberg radio telescopes \citep{2007A&A...470..539B} are used for comparison.

RM Synthesis has been introduced by \citet{Brentjens2005} for multichannel spectro-polarimetric data (data cube with two sky dimensions and wavelength) obtained by modern radio telescopes that provide observations of polarized intensity (via the Stokes parameters $Q$ and $U$) over a wide range of wavelengths $\lambda$. From the data cube RM Synthesis calculates the ``Faraday dispersion function'' $F$ (also called ``Faraday spectrum'') for each pixel on the sky plane, obtaining the Faraday cube
\begin{equation}
\label{RM}
F(\phi,l,b) = \frac{1}{\pi}\int_{\lambda_{\rm min}}^{\lambda_{\rm max}} P(\lambda^2,l,b) e^{-2i \phi \lambda^2} d\lambda^2 \, ,
\end{equation}
where $\phi$ is the Faraday depth, $l$ and $b$ are the sky coordinates of the pixel, and $\lambda_{\rm min}$ and $\lambda_{\rm max}$ define the range of available wavelengths. $F$ is measured in units of Jansky per half-width of the telescope beam and per half-width of the ``Faraday beam'' (Faraday Point Spread Function, see \citet{Brentjens2005}). We use a range of $-1000<\phi<1000\FRM$ which should be sufficient to detect all possible sources.

The Faraday depth cube generated by the transform (\ref{RM}) requires additional efforts to recognize magnetic field structures. A simple algorithm (exploited in particular by \citet{Heald2009}) is as follows. Along each line of sight the value $F^{\rm max}$ with a peak intensity at the Faraday depth $\phi^{\rm max}$ is found and stored at the corresponding pixel on the sky plane
\begin{equation}\label{maxF}
|F^{\rm max} (l, b)| = \max\limits_{\phi} |F (\phi, l, b)|  =|F (\phi^{\rm max}, l, b)| \, ,
\end{equation}
where $F^{\rm max}$ is measured in units of polarized intensity.
The maps of $|F^{\rm max}|$ and $\phi^{\rm max}$ on the sky plane $(l,b)$ of NGC~6946 are shown in Fig.~\ref{fig:data}.

\begin{figure*}
\includegraphics[width=0.3\textwidth]{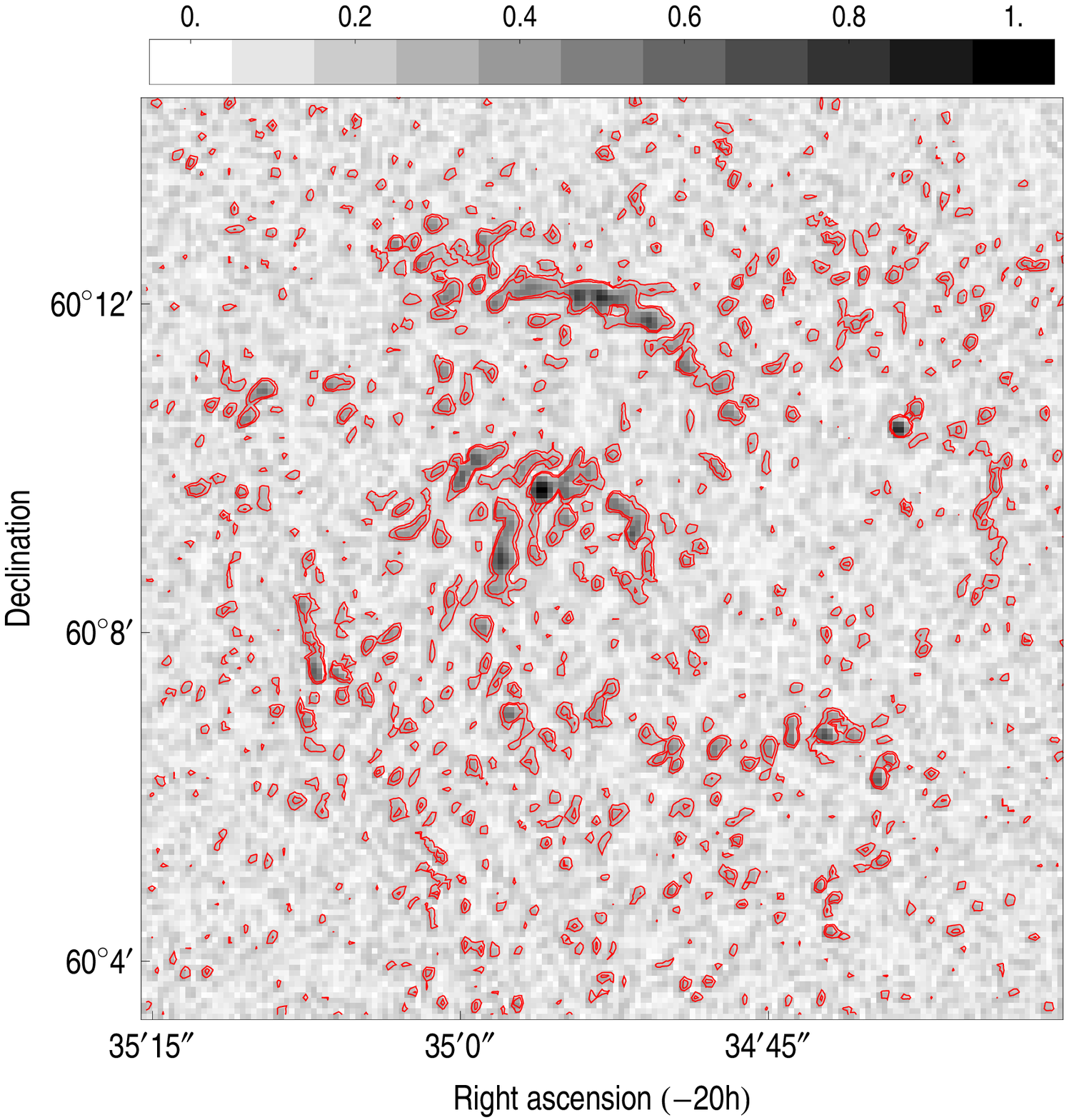}
\includegraphics[width=0.3\textwidth]{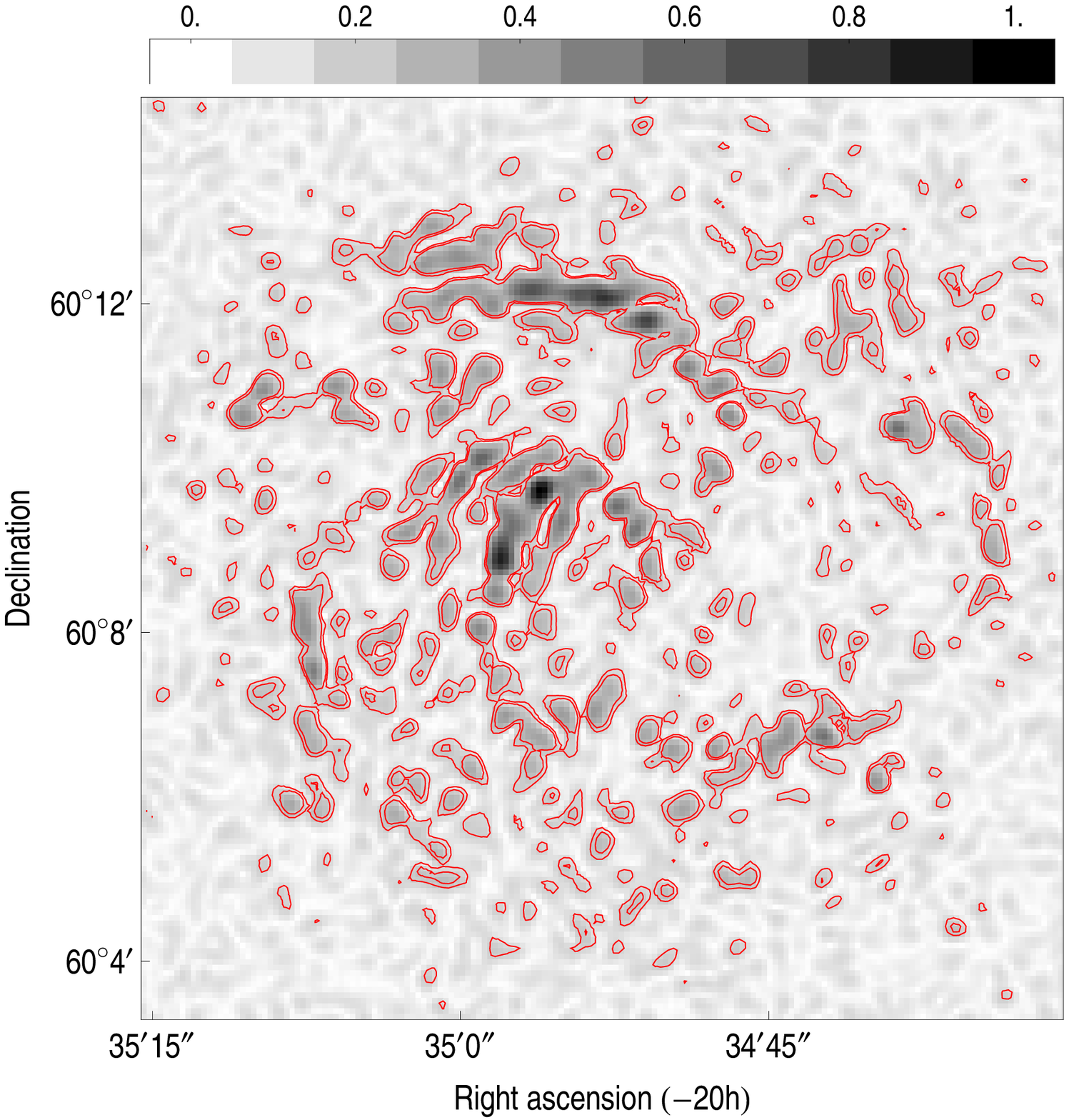}
\includegraphics[width=0.3\textwidth]{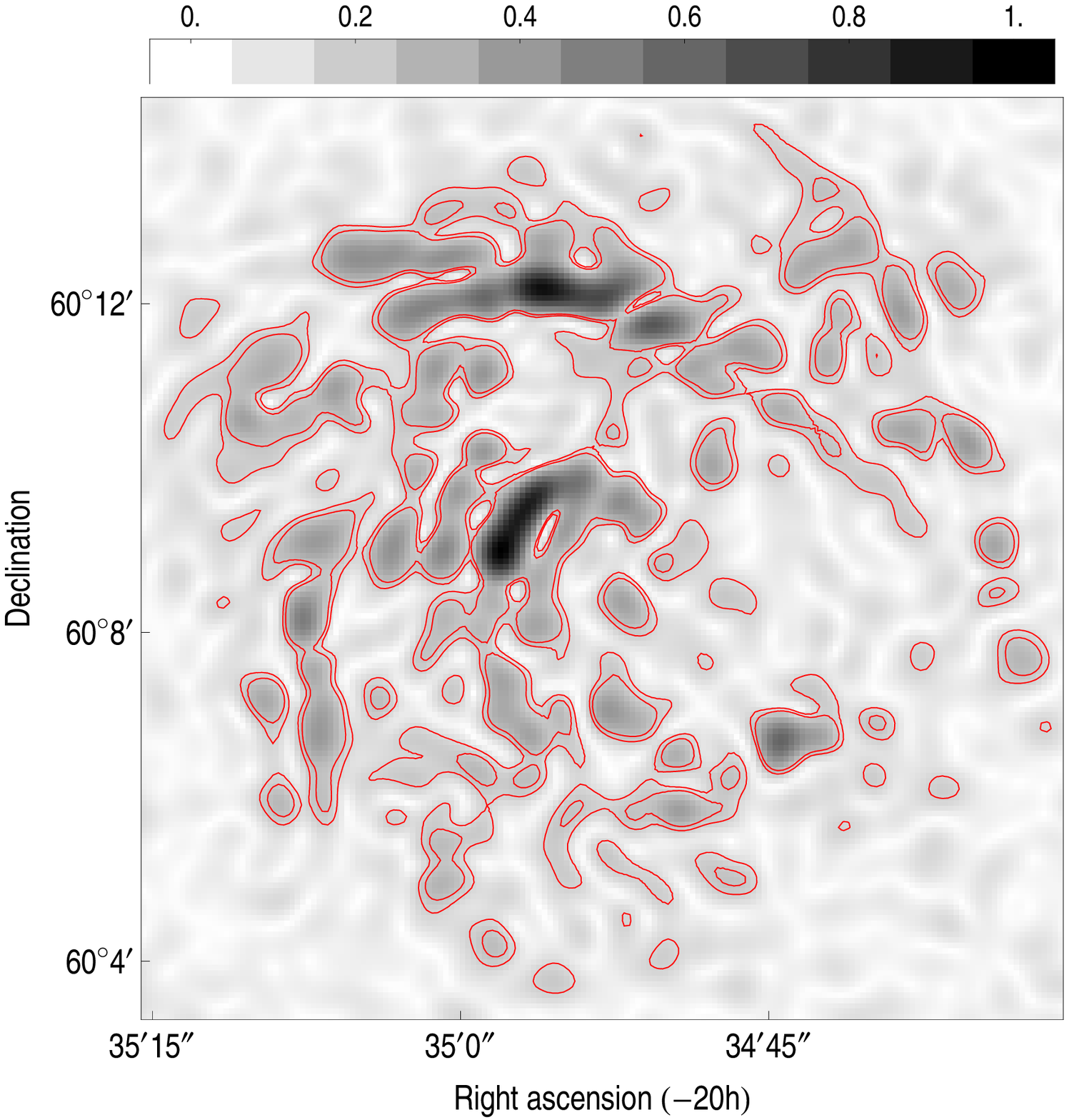}
\caption{Distribution of $|w^{40}_{a}(l,b)|$ for scales $a=8, 16, 32 \arcsec$ (from left to right) at Faraday depth $\phi=40\FRM$. The grayscale shows the normalized modulus of $w$, the red contours depict $1.5\times rms$ and $2\times rms$ levels of $w$ calculated for all values in a map.}
\label{fig:scales1}
\end{figure*}
\begin{figure*}
\includegraphics[width=0.3\textwidth]{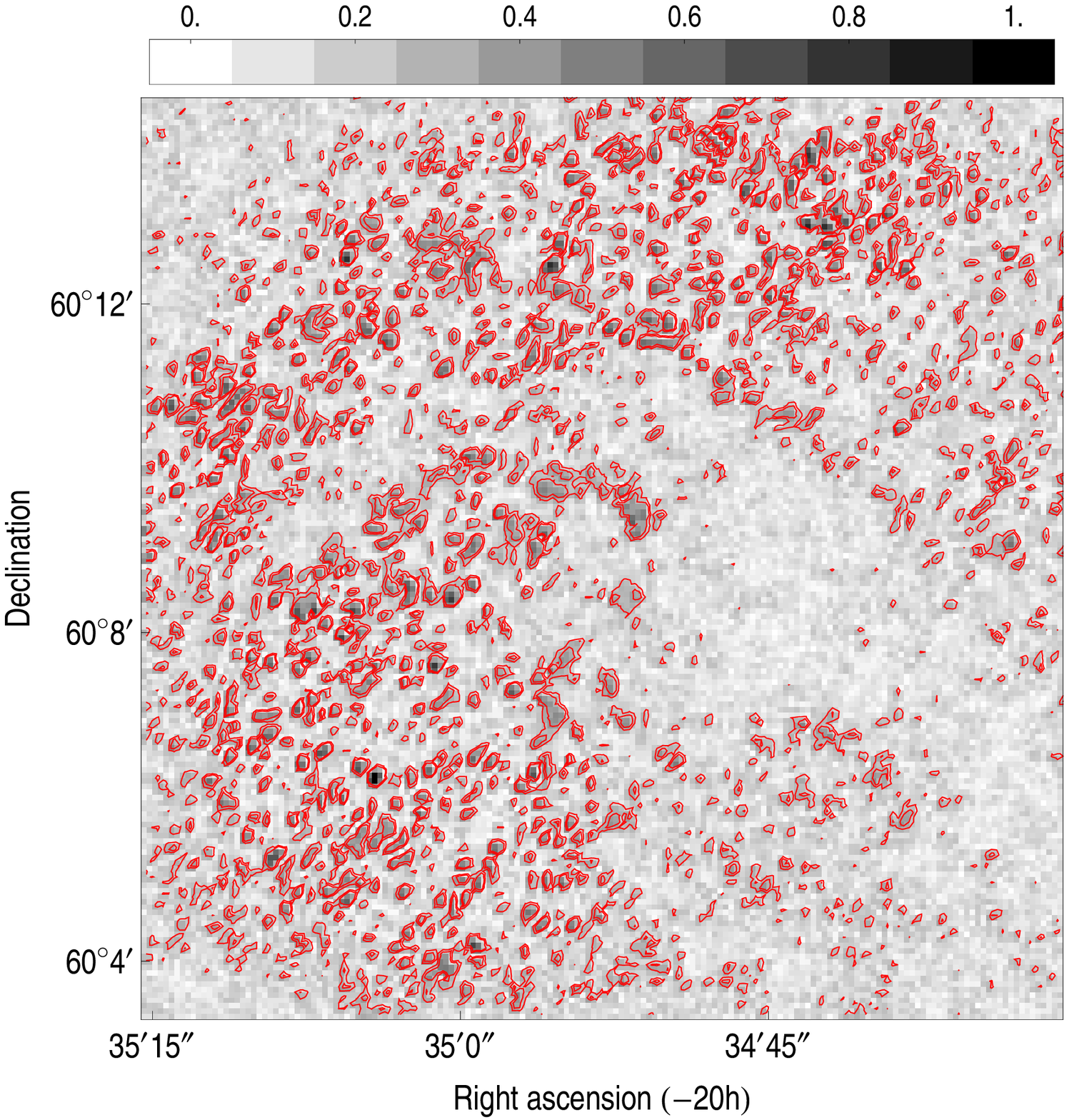}
\includegraphics[width=0.3\textwidth]{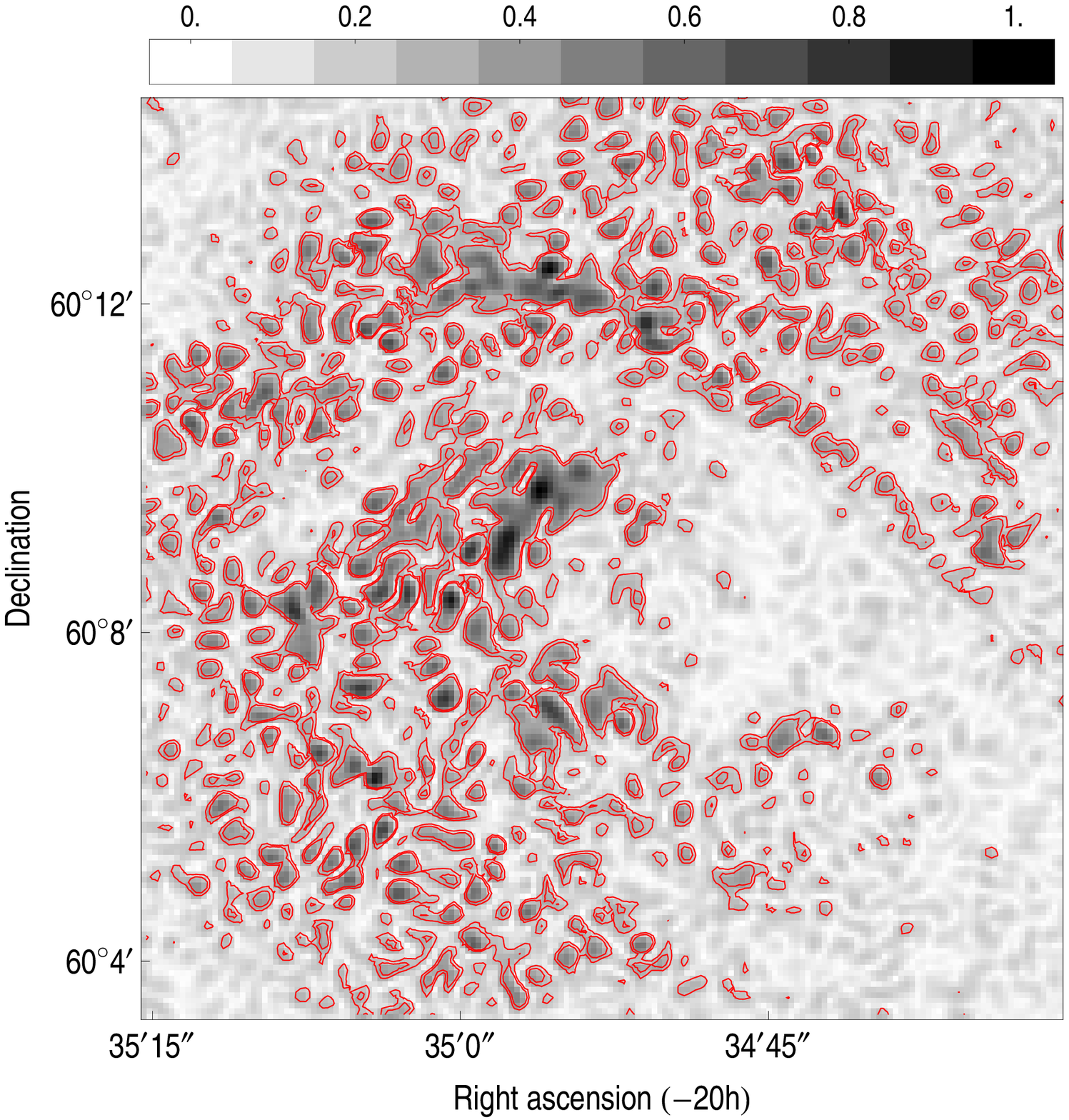}
\includegraphics[width=0.3\textwidth]{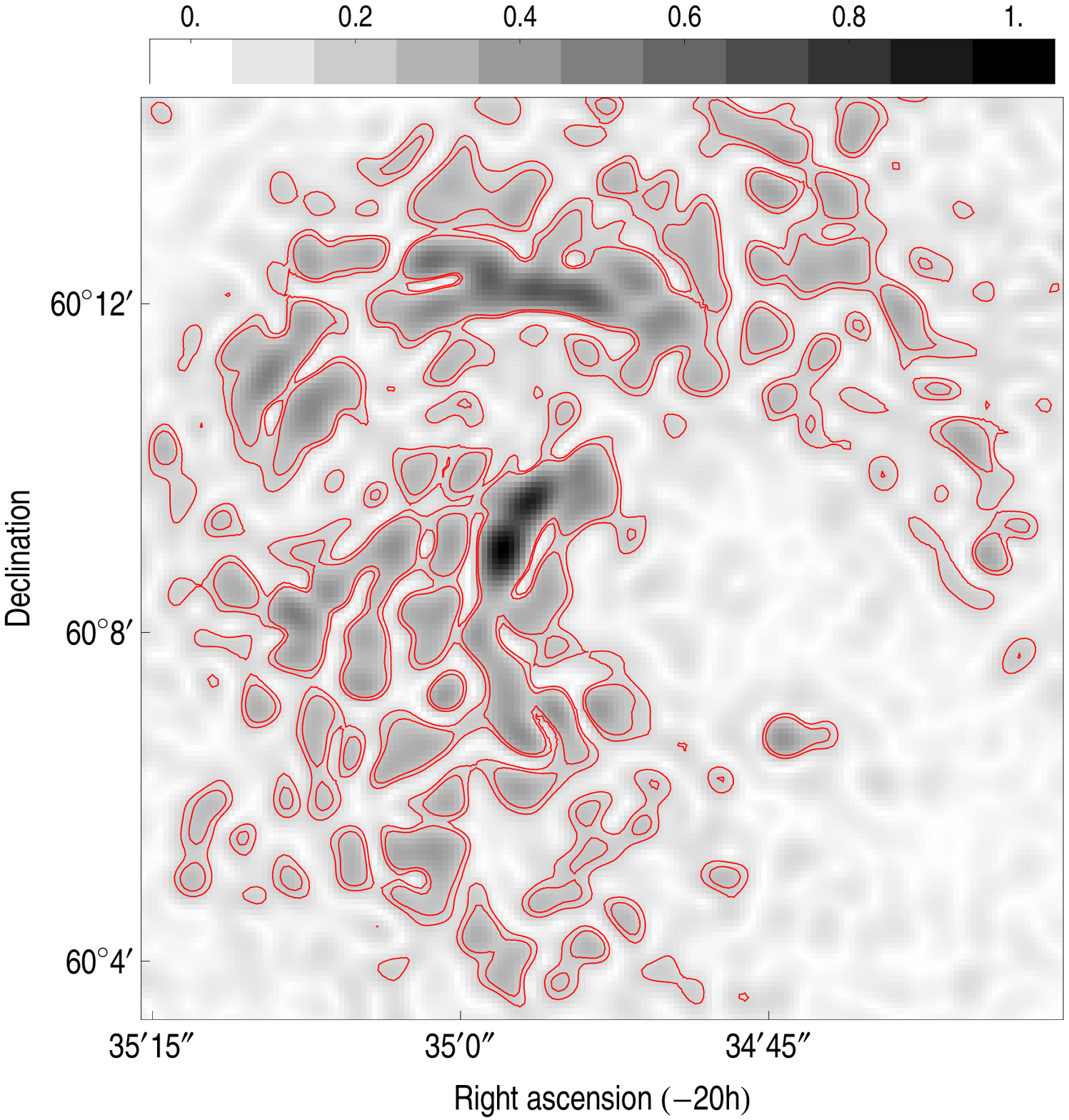}
\caption{Distribution of $|\wmaxwt_{a} (l,b)|$ for scales $a=8, 16, 32 \arcsec$ (from left to right). The grayscale shows the normalized modulus of $w$, the red contours depict $1.5\times rms$ and $2\times rms$ levels of $w$ calculated for all values in a map.}
\label{fig:scales2}
\end{figure*}
\begin{figure*}
\includegraphics[width=0.3\textwidth]{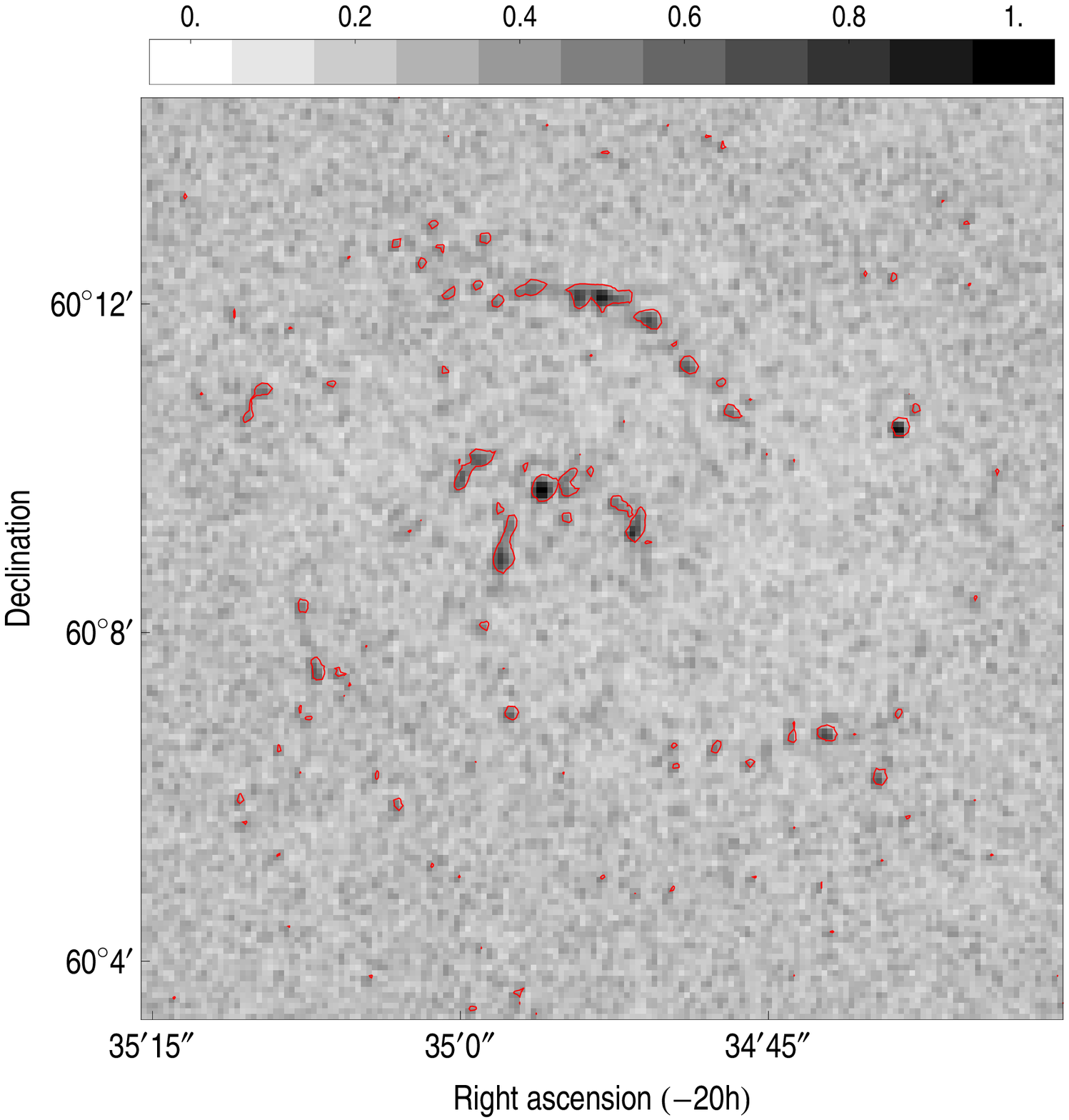}
\includegraphics[width=0.3\textwidth]{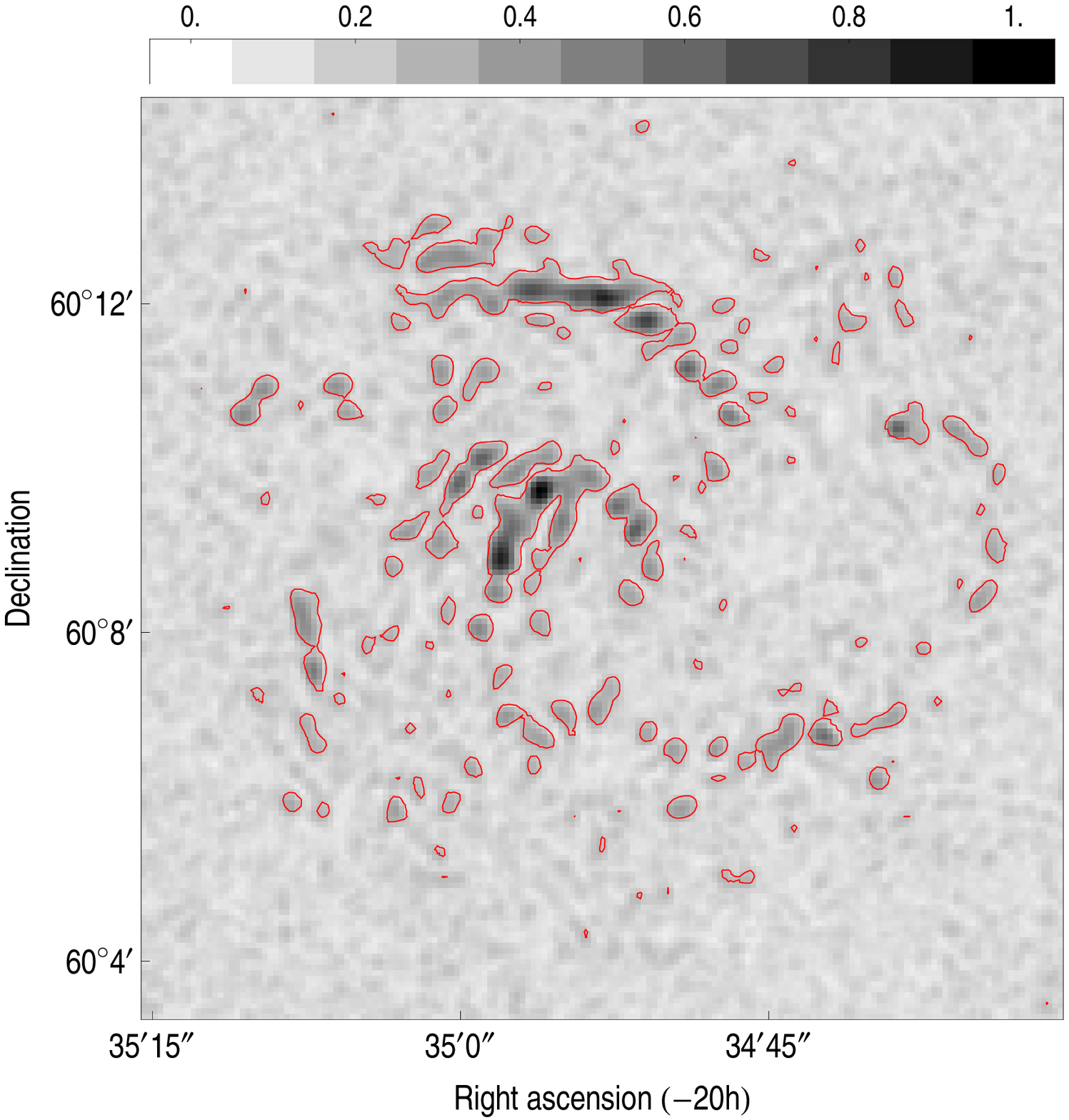}
\includegraphics[width=0.3\textwidth]{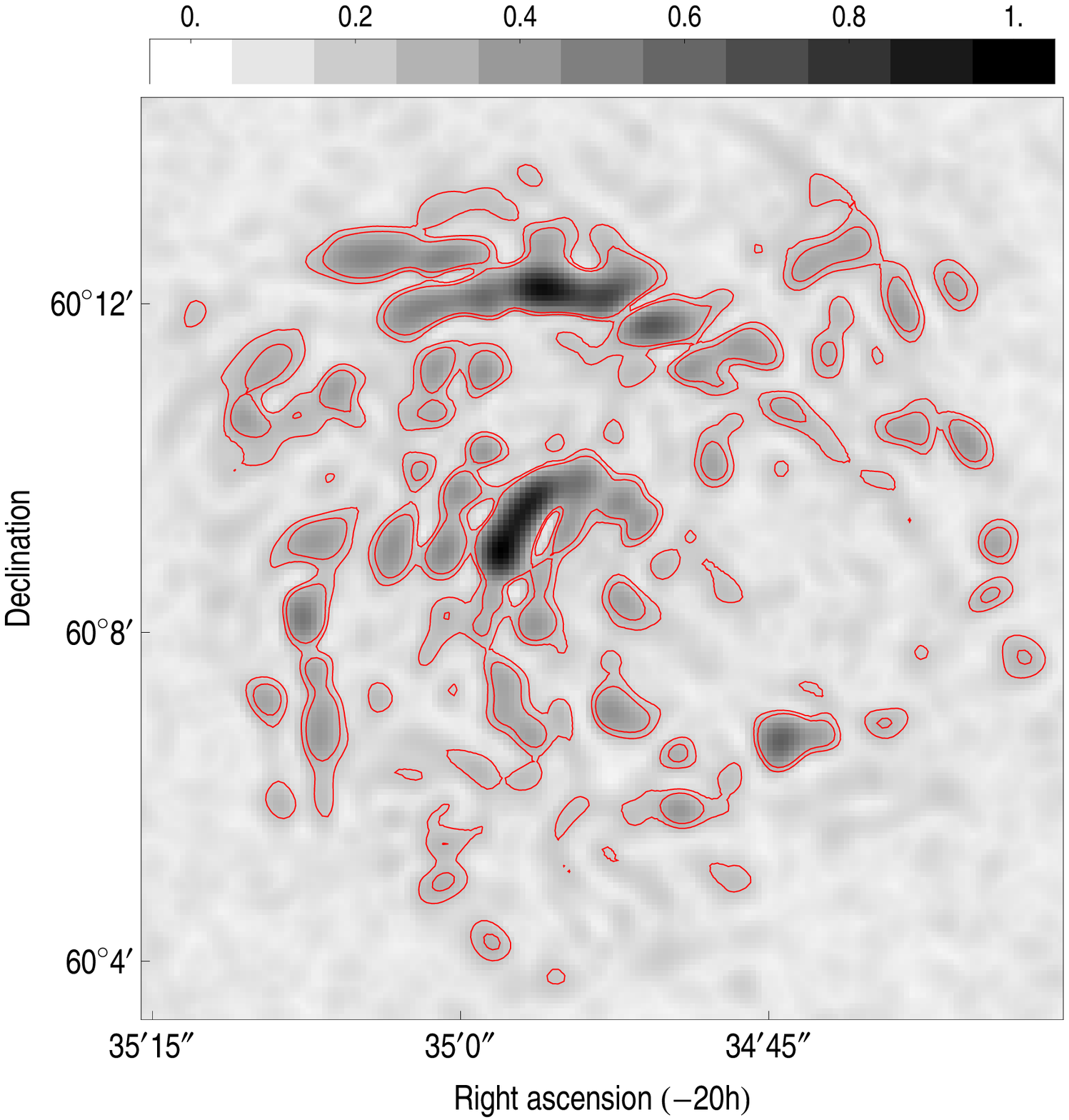}
\caption{Distribution of $|\wwtmax_a (l,b)|$ for scales $a=8, 16, 32 \arcsec$ (from left to right). The grayscale shows the normalized modulus of $w$, the red contours depict $1.5\times rms$ and $2\times rms$ levels of $w$ calculated for all values in a map.}
\label{fig:scales3}
\end{figure*}

A well-known tool for structure recognition and scale-by-scale data analysis is wavelet analysis, widely used in the interpretation of galactic images \citep[e.g.][]{Frick2001,pat2006,taba2013}. The wavelet transform can be used for optimizing RM Synthesis \citep[see][]{Frick2010,Frick2011}. In this work, wavelets are used as a spatial-scale filtering tool.

The wavelet transform of a 2D image $f(l,b)$ produces a 3D data cube (two coordinates $l$ and $b$ plus scale $a$)
\begin{multline}\label{2DW}
w_a(l,b) \equiv  W\{f(l,b)\}=\\
= \frac{1}{a^2} \infint f(l^\prime, b^\prime) \psi^*\left(\frac{l^\prime-l}{a} , \frac{b^\prime-b}{a}\right)\, dl^\prime\, db^\prime\, ,
\end{multline}
where $\psi^*$ is the complex conjugated analysing wavelet and scale $a$ is the characteristic radius of the wavelet function. $a$ is related to the full width at half power $\Theta$ as $\Theta = 2 \sqrt{2 ln 2} a \simeq 2.35 a$.

The wavelet decomposition of the Faraday spectrum for any fixed Faraday depth can be performed as
\begin{equation}\label{wav_any}
w^{\phi}_a (l,b) = W\{F(\phi,l,b)\} \, .
\end{equation}

As an example we show the wavelet decomposition applied to the map of polarized intensity at a Faraday depth of $\phi=40\FRM$ and at three different scales $a=8, 16, 32 \arcsec$ in Fig.~\ref{fig:scales1}. The Faraday depth $\phi=40\FRM$ is the average value of depth in Fig.~\ref{fig:data}, bottom.
This is the Faraday rotation of the Galactic foreground.
The patterns of ordered structures (magnetic arms) are best visible at the scale $a=16 \arcsec$, which will be used and discussed below.

In a second test, we apply the wavelet transform to the map of $F^{\rm max}$ (Fig.~\ref{fig:data}, bottom) as
\begin{equation}\label{wav_max}
\wmaxwt_a (l,b) = W\{F^{\rm max} (l,b)\} \, .
\end{equation}
The normalized modulus of $\wmaxwt_a (l,b)$  for same three scales $a=8, 16, 32 \arcsec$ is shown in Fig.~\ref{fig:scales2}. All wavelet transforms are made using an isotropic analyzing wavelet, called the Mexican hat $\psi(x,y) = (2-x^2-y^2) \, e^{-(x^2+y^2)/2}$. We conclude that the magnetic arms are not clearly present in these plots.

In general, the wavelet transform of the Faraday cube results in a 4D data array $w_a (\phi, l,b)$, which characterizes the intensity of structures of scale $a$ at Faraday depth $\phi$ at a given pixel $(l,b)$.
The key point of our approach is that we first apply the wavelet transform (\ref{wav_any}) to map at each frequency $F (\phi, l, b)$,
so we obtain one data cube per wavelet scale,
and then calculate the maximum intensity value of $w_a (\phi, l, b)$ along each line of sight (similar to $F^{\rm max}$, but for a given scale $a$)

\begin{equation}
|\wwtmax_a (l, b)| = \max\limits_{\phi} |w_a^\phi (l, b)|\, ,
\end{equation}
where the maximum is taken over the whole considered range of $\phi$.

The normalized modulus of $\wwtmax_a (l, b)$ for scales $a=8, 16, 32 \arcsec$ is shown in Fig.~\ref{fig:scales3}. At $a=16\arcsec$ (corresponding to about twice the beam size of the observations) the magnetic arms have the higher contrast in comparison with $\wmaxwt_a (l,b)$ (in Fig.~\ref{fig:scales2}). The structures enhanced in the $\wwtmax_{16\arcsec} (l, b)$ map are also visible in the $\wmaxwt_{16\arcsec} (l,b)$ map. However, the $\wmaxwt_{16\arcsec} (l,b)$ map has lot of additional structures of similar or even higher intensity that are spread over a much larger region.

The difference between the two approaches, i.e. between $\wmaxwt_a (l,b)$ and $\wwtmax_a (l, b)$, is mathematically the
inverse order of operations taking the maximum over $\phi$ and the wavelet decomposition (schematically shown in  Fig.~\ref{fig:diag}).
This difference yields additional information on the structure of large-scale magnetic field, as discussed in the next Section.

\section{Discussion and Conclusions}

The application of RM Synthesis followed by the determination of maxima of the Faraday spectra at each pixel in the sky plane at any Faraday depth does not clearly reveal any elongated arm structures (Fig.~\ref{fig:data}, top).
On the other hand, if we first apply a wavelet decomposition which isolates structures at a given scale
and then determine the maxima of the Faraday spectra at each pixel, the resulting image reveals pronounced structures that are invisible when applying RM Synthesis only.
We clearly see from Fig.~\ref{fig:scales3}, middle that the isolated structures are organized in the form of spiral arms. The elongated arms consist of a set of local maxima along the arms.

\begin{figure}
\centering
   {
    \includegraphics[width=0.38\textwidth]{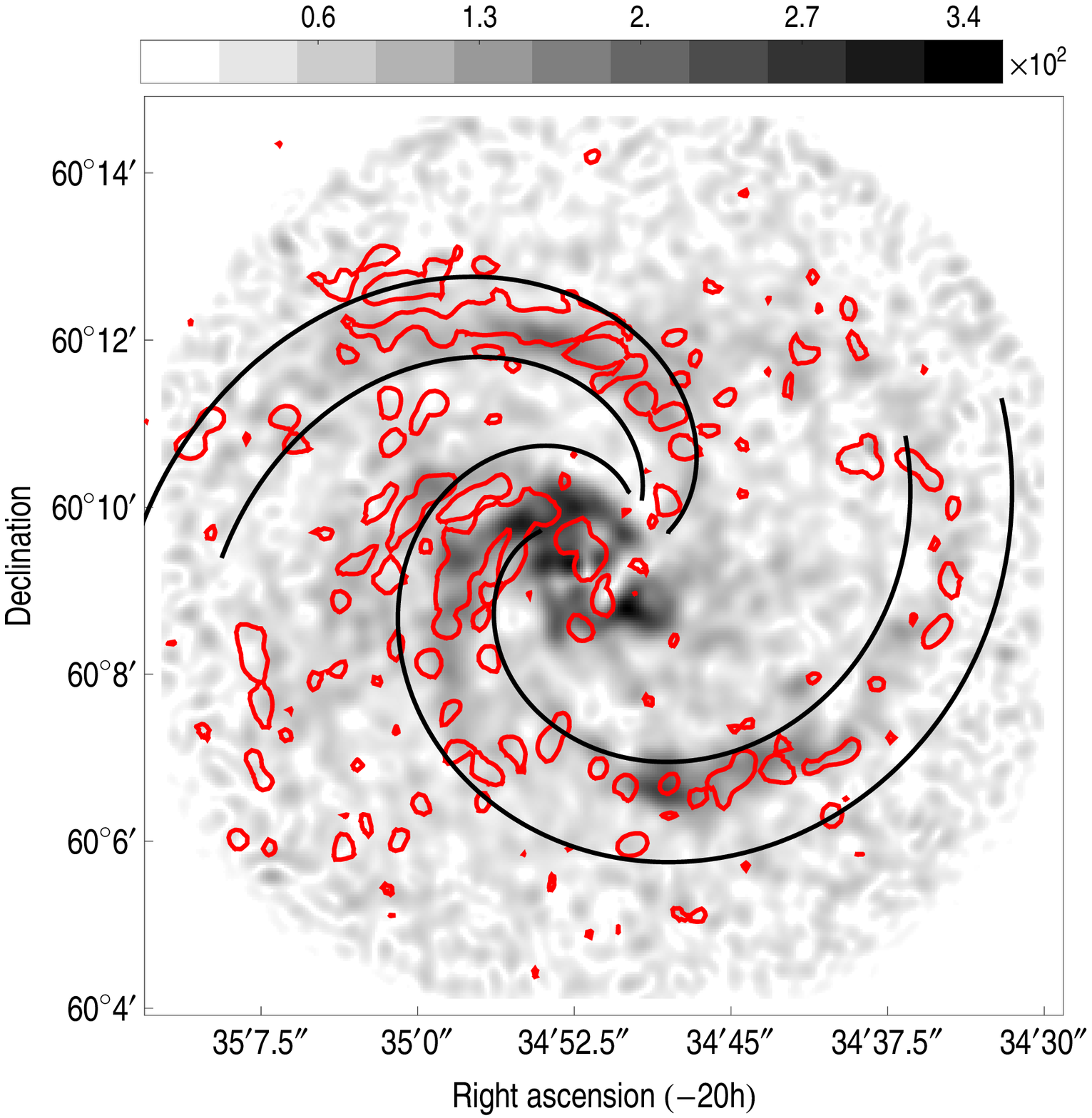}
    \includegraphics[width=0.38\textwidth]{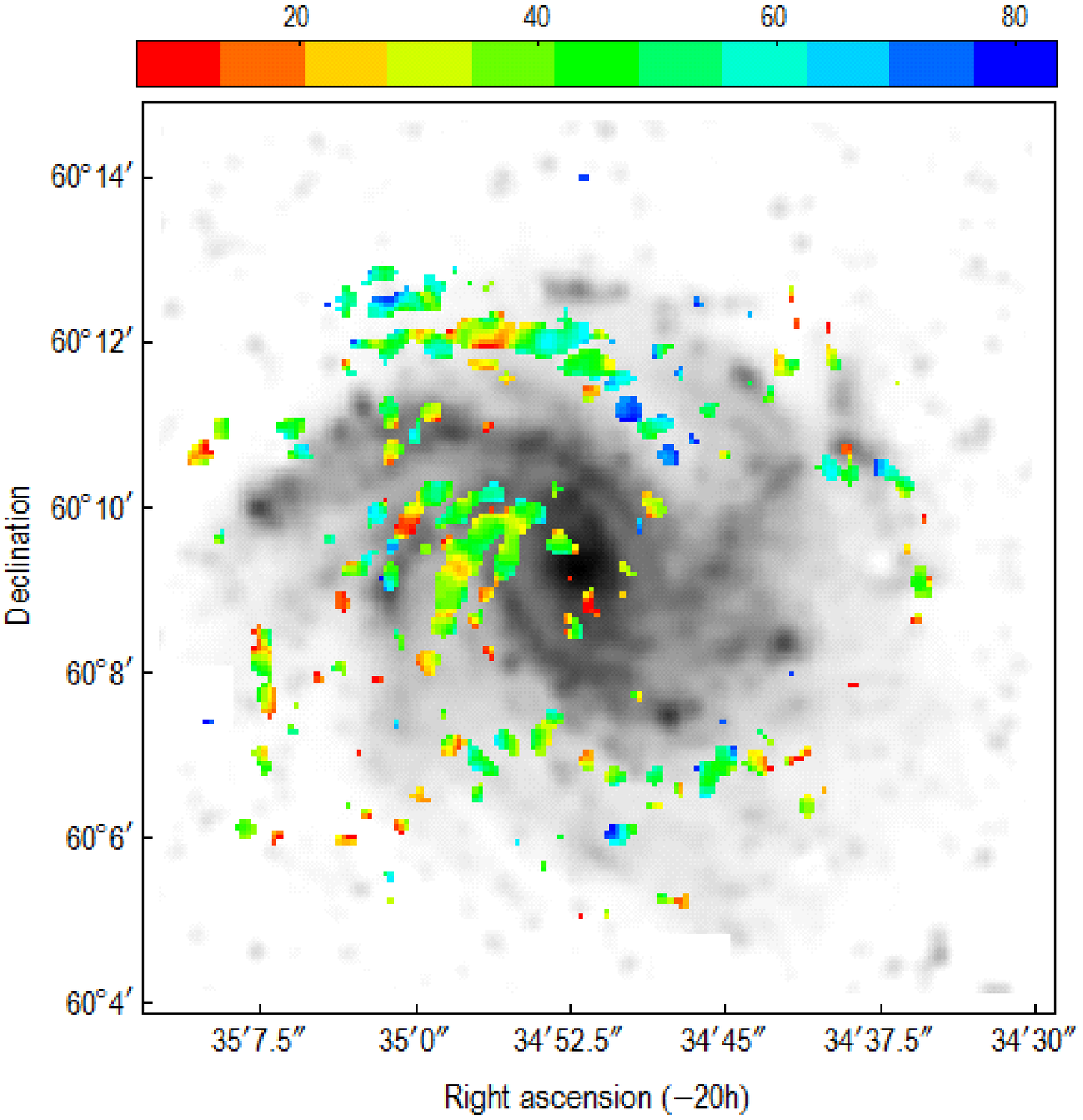}
    }
\caption{Top: Isolated magnetic arms (red contours at $1.5\times rms$) obtained from the data at 17--23\,cm with the wavelet transform for scale $a = 16\arcsec$ (Fig.~\ref{fig:scales3}, middle), and model arms for methods comparison (black contours) overlayed on the image of polarized intensity at 6\,cm wavelength (grayscale). The maximum intensity is $340\uJyb$. Bottom: Isolated magnetic arms (shown in colour depicting the Faraday depths $\phi^{\rm max}$ ($\FRM$) of $F^{\rm max}$), overlayed on an optical image (grayscale).}
\label{fig:depth}
\end{figure}

Comparing the middle plots of Fig.~\ref{fig:scales2} and Fig.~\ref{fig:scales3}, we find that $\wwtmax_a (l, b)$ has a higher contrast between arm and interarm regions than $\wmaxwt_a (l,b)$. We measure the contrast $c$  as follows. We divide the image into the magnetic arm and interarm space as it comes from 6 cm data (boundaries are shown by black contours in Fig.~\ref{fig:depth}, top) and exclude the very central part (distance from center up to 50 $\arcsec$ (equal to $1.67~\kpc$) and the outer part (more than 340~$\arcsec$ or $11.4~\kpc$). Then we calculate the intensities (root-mean-square values) of the wavelet coefficients in the arms and interarm areas and measure the contrast as the ratio between these intensities. The contrasts are given in Table~\ref{tab:rmss}. Our method enlarges this quantity by about 10\% for the northern arm and by about 15\% for the southern arm. In spite of this improvement, the contrast is still smaller than that at 6 cm.

A straightforward recommendation would be to perform observations at 6\,cm and even 3\,cm with higher sensitivity, e.g. with the SKA. As this remains unrealistic for the near future, we have to restrict our demands to longer wavelengths, e.g. with the SKA precursors MeerKAT and ASKAP. Then the suggested method is able to improve the contrast up to 15\%, which can be sufficient to isolate the arms in the images. We remind that we used NGC~6946 as an illustrative example because the position of magnetic arms is known from 6\,cm data. Applications of the method are recommended for galaxies where 6\,cm data are absent.

\begin{table}
\caption{Comparison of contrast for various distributions between arm and interarm regions for wavelet coefficients at scale $a=16\arcsec\approx 535\pc$, versus 6 cm. Errors were estimated by standard deviation at 30\% bootstrapping of the sets}
\label{tab:rmss}
\begin{tabular}{cccc}
\hline
             &  $\wmaxwt_a $      & $\wwtmax_a$        & 6 cm              \\
             \hline
north. arm &  $1.398 \pm 0.007$ & $1.517 \pm 0.007$  & $2.002 \pm 0.009$ \\
south. arm &  $1.116 \pm 0.005$ & $1.273 \pm 0.005$  & $2.310 \pm 0.010$ \\
both arms    &  $1.212 \pm 0.005$ & $1.354 \pm 0.004$  & $2.221 \pm 0.008$ \\
\hline
\end{tabular}
\end{table}

Structures of $\wwtmax_a (l, b)$ fit to the large-scale structures that are visible in the polarization map at 6$\cm$ wavelength (see Fig.~\ref{fig:depth}, top), located between the optical arms (see Fig.~\ref{fig:depth}, bottom). We obtain enlargement of contrast for small $a$ only. It means that the method is sensitive to small-scale details in the image.
These details  can correspond to real small-scale structures of the magnetic field or be the result of Faraday effects on the polarized emission from the large-scale magnetic field, to be figured out with numerical tests
which is presented below. Note that the polarized intensity map at 17-23$\cm$ (see Fig.~\ref{fig:data}, top) revels some detail of magnetic arms which are identified in Fig.~\ref{fig:depth}. The point is however that the details in  Fig.~\ref{fig:data}, top are embedded in diffuse surrounding and its relation with magnetic arms remains unclear.

To illustrate the idea of the method we constructed an artificial example, producing a data cube for the same set of channels as in data for NGC~6946 analysed before. We simulate the polarized emission (in the computational box 25$\times$25$\times$5$\kpc$)
emerging from a large-scale magnetic field in the galactic disk
observed face-on that is embedded in a 3D homogeneous isotropic turbulent field  in the halo. The large-scale field has only an azimuthal component perpendicular to the line of sight and Gaussian shape
\begin{equation}
B_\varphi(r,\varphi,z)=B_0 \exp{\left[-\left(\frac{r}{r_0}\right)^2-\left(\frac{z}{z_0}\right)^2\right]} \, ,
\end{equation}
where the radial Gaussian scalelength is $r_0=10$ $\kpc$, the vertical Gaussian scale is $z_0=1$ $\kpc$ and the strength of the large-scale field is $B_0=3$ $\mu$G.
The homogeneous turbulent field was simulated by a numerical model that allows to control the spectral law and the characteristic scale of turbulence $l_t$ (providing the maximum of energy spectrum) \citep{Stepanov2014}. We do not consider any possible inhomogeneity of the turbulent field in the midplane. We choose a spectral slope of -5/3 at smaller scales than $l_t$ and +2 at larger scales.
We choose three different values of $l_t=100, 200, 400$ $\pc$ in order to search for a possible dependence.

The rms strength of the turbulent field is taken as 1\,$\mu$G.
The number densities of relativistic and thermal electrons are assumed to be constant, namely $n_c=1$ cm$^{-3}$
and $n_e=0.1$ cm$^{-3}$.
The large-scale disk magnetic field corresponds to a Faraday-thin source of the synchrotron emission
which does not cause significant Faraday depolarization. The turbulent field acts as a Faraday screen that does not contribute much to the emission but depolarizes it and disperses it to different Faraday depth. The output of the model is two data cubes of Q and U with coordinates and frequency.

\begin{figure}
\centering
\includegraphics[width=0.4\textwidth]{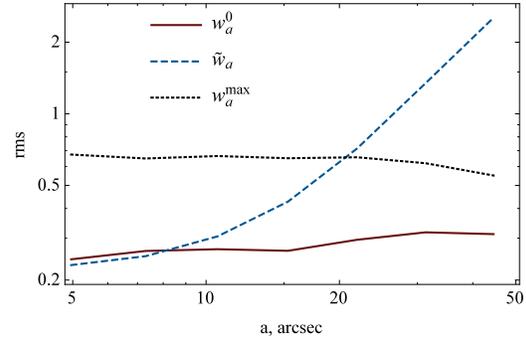}
\caption{The root-mean-square spectral power as a function of scale $a$ for the simulated data: thick line -- $w^{40}_a$, dashed  line -- $\wmaxwt_a$, dotted line -- $\wwtmax_a$. The characteristic scale of turbulence is $l_t=72\arcsec$, corresponding to 200 $\pc$. The noise level of the synthetic signal is zero. }
\label{fig:test4spec}
\end{figure}

Next, we perform RM Synthesis on the artificial data cube using the same wavelength range as in the case of NGC~6946 and
 calculate the wavelet coefficients $\wmaxwt_a (l,b)$ and $\wwtmax_a (l, b)$.
The wavelet filter doesn't influence on Faraday spectra directly, however, it suppresses peaks in 2D map whose scales are different from scale $a$.
The intensities (spectral power) of $\wmaxwt_a (l,b)$ and $\wwtmax_a (l, b)$
versus scale $a$ are shown in Fig.~\ref{fig:test4spec}.

The intensity of $w^{40}_a$ is on a low level on all scales because the polarized intensity at Faraday depth $\phi=40\FRM$ does not contain spectral power at large scales.
The intensity of $\wmaxwt_a (l,b)$ substantially increases with scale because only large scales are prominent, while the intensity of $\wwtmax_a (l, b)$ is practically constant, it detects power on all scales.
If the large scales are well represented in the data (i.e. if there is no strong Faraday depolarization and most spectral power is on large scales), the standard method of structure recognition ($\wmaxwt_a (l,b)$) works best and reveals the structure of the large-scale magnetic field. If, however, Faraday depolarization is strong and there is no chance to recognize the large scales directly, our new method opens an additional possibility to find imprints of the large scales at small scales; the small scales are much better recognized by $\wwtmax_a (l, b)$ compared to $\wmaxwt_a (l,b)$.
Late application of wavelet transform in $\wwtmax_a (l, b)$ suppresses the regions in the extended disk but keeps the regions in the magnetic arms, because the turbulent field is weaker there and hence there is less Faraday dispersion, so that the structures are less randomized.
The scale of the crossing point in Fig.~\ref{fig:test4spec} depends of the properties of turbulence and observation wavelength and gives the upper limit in scale  below which the suggested technique is suitable.

\begin{figure}
\centering
    \includegraphics[width=0.25\textwidth]{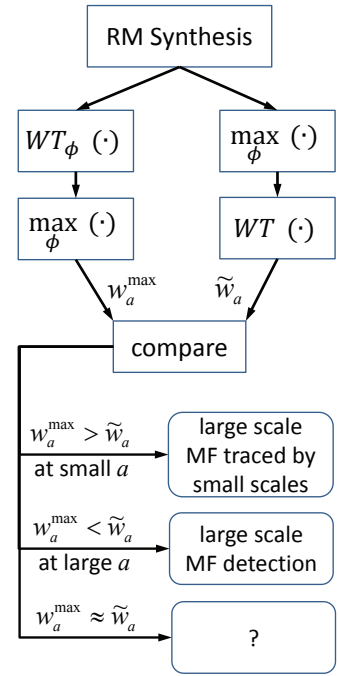}
\caption{Scheme of a heuristic approach. Notation $(\cdot)$ means an operator argument taken as a result from the previous step.}
\label{fig:diag}
\end{figure}

We checked the minimal requirement of recovering the Faraday spectrum from the observational frequency band. The FWHM of the Faraday Point Spread Function (FPSF), or equivalently, the resolution in Faraday depth space $\triangle\phi_{\rm FPSF}$ should be comparable at least to the rms dispersion of Faraday rotation caused by the turbulent magnetic field. Following \citet{Brentjens2005} the estimate
$$\Delta\phi_{\rm FPSF}=\frac{2 \sqrt{3}}{\lambda^2_{\rm max}-\lambda^2_{\rm min}}$$
gives $\Delta\phi_{\rm FPSF}\approx170 \FRM $  for the observations used here. This is larger than the dispersion of Faraday rotation (about 40 $\FRM$)
in Fig.~\ref{fig:data} (bottom). It explains why enlargement of contrast obtained by our method (see Table~\ref{tab:rmss}) remains moderate however sufficient to isolate magnetic arms in Fig.~\ref{fig:scales3}.
If $\Delta\phi_{\rm FPSF}$ is very large, then $\phi^{\rm max}$ is about the same for all lines of sight, so that $\wmaxwt_a \approx \wwtmax_a$ and our method give  the same results as the traditional one.

Our model is admittedly simplistic. Observations at longer wavelengths generally probe magnetic structures that are farther from the galaxy midplane and hence close to the observer \citep[e.g.][]{2010A&A...514A..42B}. Presuming that the magnetic field morphology has some vertical structure, changes in the large-scale morphological features are expected when observing at progressively lower and lower frequencies. If the large-scale fields in the magnetic arms are tied to
the star-forming ISM, we may expect to see them vanish at lower frequencies, where mainly the thick disk or halo is observed rather than emission from the disk that is depolarized by star-formation induced turbulence.
In the case of NGC~6946, our result shows that the magnetic arms extend sufficiently high into the thick disk or halo, so that their imprints can still be detected at wavelengths around 20\,cm.

The overall scheme of our analysis is shown in Fig.~\ref{fig:diag}.
In summary, the distribution of small-scale magnetic fields as recognized by wavelet filtering of spatial scales traces the locations of the large-scale field (e.g. in the magnetic arms), if the imprints of large-scale fields are randomized by Faraday depolarization.
This method is a powerful tool to analyze spectro-polarimetric data cubes obtained at long radio wavelengths.
It should be applied to further galaxies from the survey by \citet{Heald2009} and to galaxies observed with the VLA in L-band (1--2\,GHz) and with new-generation radio telescopes like LOFAR, ASKAP, MeerKAT and SKA.

\acknowledgements
The authors thank local and anonimous referees for substantial questions which are significantly improves the understandability and conclusiveness of the paper.
Rainer Beck acknowledges support by DFG Research Unit FOR1254. Numerical simulations were performed on the supercomputers URAN and TRITON of Russian Academy of Science, Ural Branch.

\bibliographystyle{an}
\bibliography{lit}

\begin{thebibliography}{21}
\expandafter\ifx\csname natexlab\endcsname\relax\def\natexlab#1{#1}\fi

\bibitem[{{Beck}(2007)}]{2007A&A...470..539B}
{Beck}, R. 2007, \aap, 470, 539

\bibitem[{{Beck}(2015)}]{Beck2015}
{Beck}, R. 2015, \aap, 578, A93

\bibitem[{{Beck} {et~al.}(1996){Beck}, {Brandenburg}, {Moss}, {Shukurov}, \&
  {Sokoloff}}]{Betal96}
{Beck}, R., {Brandenburg}, A., {Moss}, D., {Shukurov}, A., \& {Sokoloff}, D.
  1996, \araa, 34, 155

\bibitem[{{Beck} {et~al.}(2012){Beck}, {Frick}, {Stepanov}, \&
  {Sokoloff}}]{Beck2012}
{Beck}, R., {Frick}, P., {Stepanov}, R., \& {Sokoloff}, D. 2012, \aap, 543,
  A113

\bibitem[{{Braun} {et~al.}(2010){Braun}, {Heald}, \&
  {Beck}}]{2010A&A...514A..42B}
{Braun}, R., {Heald}, G., \& {Beck}, R. 2010, \aap, 514, A42

\bibitem[{Brentjens \& de~Bruyn(2005)}]{Brentjens2005}
Brentjens, M.~A. \& de~Bruyn, A.~G. 2005, \aap, 441, 1217

\bibitem[{{Burn}(1966)}]{B66}
{Burn}, B.~J. 1966, \mnras, 133, 67

\bibitem[{{Chamandy} {et~al.}(2013){Chamandy}, {Subramanian}, \&
  {Shukurov}}]{Anvetal13}
{Chamandy}, L., {Subramanian}, K., \& {Shukurov}, A. 2013, \mnras, 428, 3569

\bibitem[{{Frick} {et~al.}(2001){Frick}, {Beck}, {Berkhuijsen}, \&
  {Patrickeyev}}]{Frick2001}
{Frick}, P., {Beck}, R., {Berkhuijsen}, E.~M., \& {Patrickeyev}, I. 2001,
  \mnras, 327, 1145

\bibitem[{{Frick} {et~al.}(2010){Frick}, {Sokoloff}, {Stepanov}, \&
  {Beck}}]{Frick2010}
{Frick}, P., {Sokoloff}, D., {Stepanov}, R., \& {Beck}, R. 2010, \mnras, 401,
  L24

\bibitem[{{Frick} {et~al.}(2011){Frick}, {Sokoloff}, {Stepanov}, \&
  {Beck}}]{Frick2011}
{Frick}, P., {Sokoloff}, D., {Stepanov}, R., \& {Beck}, R. 2011, \mnras, 414,
  2540

\bibitem[{{Frick} {et~al.}(2016){Frick}, {Stepanov}, {Beck}, {Sokoloff},
  {Shukurov}, {Ehle}, \& {Lundgren}}]{Frick16}
{Frick}, P., {Stepanov}, R., {Beck}, R., {et~al.} 2016, \aap, 585, A21

\bibitem[{{Heald} {et~al.}(2009){Heald}, {Braun}, \& {Edmonds}}]{Heald2009}
{Heald}, G., {Braun}, R., \& {Edmonds}, R. 2009, \aap, 503, 409

\bibitem[{{La Porta} \& {Burigana}(2006)}]{2006A&A...457....1L}
{La Porta}, L. \& {Burigana}, C. 2006, \aap, 457, 1

\bibitem[{{Moss} {et~al.}(2015){Moss}, {Stepanov}, {Krause}, {Beck}, \&
  {Sokoloff}}]{Metal15}
{Moss}, D., {Stepanov}, R., {Krause}, M., {Beck}, R., \& {Sokoloff}, D. 2015,
  \aap, 578, A94

\bibitem[{{Patrikeev} {et~al.}(2006){Patrikeev}, {Fletcher}, {Stepanov},
  {Beck}, {Berkhuijsen}, {Frick}, \& {Horellou}}]{pat2006}
{Patrikeev}, I., {Fletcher}, A., {Stepanov}, R., {et~al.} 2006, \aap, 458, 441

\bibitem[{{Pshirkov} {et~al.}(2011){Pshirkov}, {Tinyakov}, {Kronberg}, \&
  {Newton-McGee}}]{Pr}
{Pshirkov}, M.~S., {Tinyakov}, P.~G., {Kronberg}, P.~P., \& {Newton-McGee},
  K.~J. 2011, \apj, 738, 192

\bibitem[{{Ruzmaikin} \& {Sokoloff}(1979)}]{RS79}
{Ruzmaikin}, A.~A. \& {Sokoloff}, D.~D. 1979, \aap, 78, 1

\bibitem[{{Sokoloff} {et~al.}(1998){Sokoloff}, {Bykov}, {Shukurov},
  {Berkhuijsen}, {Beck}, \& {Poezd}}]{Setal98}
{Sokoloff}, D.~D., {Bykov}, A.~A., {Shukurov}, A., {et~al.} 1998, \mnras, 299,
  189

\bibitem[{{Stepanov} {et~al.}(2014){Stepanov}, {Shukurov}, {Fletcher}, {Beck},
  {La Porta}, \& {Tabatabaei}}]{Stepanov2014}
{Stepanov}, R., {Shukurov}, A., {Fletcher}, A., {et~al.} 2014, \mnras, 437,
  2201

\bibitem[{{Tabatabaei} {et~al.}(2013){Tabatabaei}, {Berkhuijsen}, {Frick},
  {Beck}, \& {Schinnerer}}]{taba2013}
{Tabatabaei}, F.~S., {Berkhuijsen}, E.~M., {Frick}, P., {Beck}, R., \&
  {Schinnerer}, E. 2013, \aap, 557, A129

\end{thebibliography}



\end{document}